\documentclass{article}
\usepackage{cite}

\begin{document}

\date{}
\title{On the Q-BOR-FDTD method}
\author{Francisco M. Fern\'{a}ndez\thanks{%
fernande@quimica.unlp.edu.ar} \\
INIFTA, DQT, Sucursal 4, C. C. 16, \\
1900 La Plata, Argentina}
\maketitle

\begin{abstract}
We discuss a recently proposed approach termed Q-BOR-FDTD method and develop
its main equations in a clearer and more rigorous way. We show that it is
unsuitable for the calculation of the eigenfunctions in the case of
degenerate states and propose an improvement to overcome such limitation.
\end{abstract}

\section{Introduction}

\label{sec:intro}

In a recent paper Firoozi et al\cite{FMKJ22} proposed a numerical approach,
named quantum body of revolution finite-difference time-domain (Q-BOR-FDTD)
method, for the calculation of eigenvalues and eigenfunctions of the
Schr\"{o}dinger equation. They applied it to a variety of simple models
(some of them rather trivial) like the particle in a spherical box, the
three-dimensional harmonic oscillator, the particle in a cylindrical box, a
cone-like quantum dot and the spherical quantum dot with hydrogenic
impurity. They obtained some of the lowest eigenvalues and eigenfunctions
for each of them.

The purpose of this paper is a somewhat more rigorous discussion of the
approach. In section~\ref{sec:method} we outline the main ideas about the
method; in section~\ref{sec:examples} we disclose a deficiency of the method
by means of one of the examples and put forward an improvement; finally, in
section~\ref{sec:conclusions} we summarize the main results and draw
conclusions.

\section{The Method}

\label{sec:method}

The purpose of this section is a more rigorous discussion of the Q-BOR-FDTD
method with the purpose of disclosing some of the limitations overlooked by
Firoozi et al\cite{FMKJ22}.

In order to obtain the eigenvalues $E_{n}$ and eigenfunctions $\varphi _{n}$
of a given Hamiltonian operator $H$
\begin{equation}
H\varphi _{n}=E_{n}\varphi _{n},\;n=0,1,\ldots ,  \label{eq:Schro_time_indep}
\end{equation}
Firoozi et al proposed an approach based on the equation
\begin{equation}
\frac{\partial \psi (\mathbf{r},\tau )}{\partial \tau }=-H\psi (\mathbf{r}%
,\tau ).  \label{eq:Schro_time_dep}
\end{equation}
If
\begin{equation}
\psi (\mathbf{r},0)=\chi =\sum_{j=0}^{\infty }c_{j}\varphi
_{j},\;\left\langle \varphi _{i}\right| \left. \varphi _{j}\right\rangle
=\delta _{ij},  \label{eq:chi}
\end{equation}
then, the solution to equation (\ref{eq:Schro_time_dep}) is
\begin{equation}
\psi (\mathbf{r},\tau )=e^{-\tau H}\chi =\sum_{j=0}^{\infty }c_{j}\varphi
_{j}e^{-\tau E_{j}},  \label{eq:psi(r,tau)}
\end{equation}
The main idea behind the approach is that
\begin{equation}
\lim\limits_{\tau \rightarrow \infty }\left\langle H\right\rangle
=E_{0},\;\left\langle H\right\rangle =\frac{\left\langle \psi \right|
H\left| \psi \right\rangle }{\left\langle \psi \right| \left. \psi
\right\rangle }.  \label{eq:<H>_E0}
\end{equation}
Therefore, it is \textit{only} necessary to solve equation (\ref
{eq:Schro_time_dep}) for sufficiently large values of $\tau $. To this end
Firoozi et al\cite{FMKJ22} resorted to a straightforward discretization
algorithm that produces $\psi (\mathbf{r}_{i},\tau _{j})$ at the points of a
suitably chosen mesh. Since
\begin{equation}
\frac{\partial \left\langle H\right\rangle }{\partial \tau }=2\left(
\left\langle H\right\rangle ^{2}-\left\langle H^{2}\right\rangle \right) <0,
\label{eq:d<H>/dtau}
\end{equation}
we conclude that the numerical eigenvalue should approach the exact one from
above as $\tau $ increases. This expression, which was not mentioned by
Firoozi et al, is a useful test for the accuracy of the calculation.

In order to obtain the ground-state eigenfunction Firoozi et al\cite{FMKJ22}
resorted to the obvious expression
\begin{equation}
\tilde{\varphi}_{0}=\lim\limits_{\tau \rightarrow \infty }\frac{\psi }{\sqrt{%
\left\langle \psi \right| \left. \psi \right\rangle }}=\frac{c_{0}}{\left|
c_{0}\right| }\varphi _{0},  \label{eq:varphi_0}
\end{equation}
that they wrote in a rather imprecise way. We can extend this procedure to
excited states quite easily if we take into account that
\begin{equation}
\psi _{1}=\left( 1-\left| \tilde{\varphi}_{0}\right\rangle \left\langle
\tilde{\varphi}_{0}\right| \right) \psi =\sum_{j=1}^{\infty }c_{j}\varphi
_{j}e^{-\tau E_{j}},  \label{eq:psi_1}
\end{equation}
that Firoozi et al also expressed unclearly. Therefore, we can repeat the
procedure outlined above and obtain $E_{1}$ and $\tilde{\varphi}_{1}$ and
then proceed to other excited states. Obviously, we have tacitly assumed
that there are no degenerate states; that is to say: $E_{j}<E_{j+1}$, $%
j=0,1,\ldots $. Firoozi et al\cite{FMKJ22} did not pay attention to this
point and applied the approach to problems with degenerate states.

Suppose that
\begin{equation}
H\varphi _{n,i}=E_{n}\varphi _{n,i},\;n=0,1,\ldots ,\;i=1,2,\ldots
,g_{n},\;\left\langle \varphi _{n,i}\right| \left. \varphi _{n^{\prime
},i^{\prime }}\right\rangle =\delta _{nn^{\prime }}\delta _{ii^{\prime }}.
\label{eq:Schro_time_indep_degenerate}
\end{equation}
In this case we have
\begin{equation}
\chi =\sum_{n=0}^{\infty }\sum_{i=0}^{g_{n}}c_{n,i}\varphi _{n,i},
\label{eq:chi_degenerate}
\end{equation}
and
\begin{equation}
\psi (\mathbf{r},\tau )=\sum_{n=0}^{\infty }e^{-\tau
E_{n}}\sum_{i=0}^{g_{n}}c_{n,i}\varphi _{n,i}.
\label{eq:psi_expansion_degenerate}
\end{equation}
It is convenient to define
\begin{equation}
u_{n}=\frac{\sum_{i=0}^{g_{n}}c_{n,i}\varphi _{n,i}}{\sqrt{%
\sum_{i=0}^{g_{n}}\left| c_{n,i}\right| ^{2}}},\;\left\langle u_{i}\right|
\left. u_{j}\right\rangle =\delta _{ij},  \label{eq:u_n}
\end{equation}
and rewrite equation (\ref{eq:psi_expansion_degenerate}) as
\begin{equation}
\psi (\mathbf{r},\tau )=\sum_{n=0}^{\infty }a_{n}u_{n}e^{-\tau
E_{n}},\;a_{n}=\sqrt{\sum_{i=0}^{g_{n}}\left| c_{n,i}\right| ^{2}}.
\label{eq:psi_expansion_degenerate_2}
\end{equation}
We appreciate that the approach outlined above provides the eigenvalues but
not the eigenfunctions. Instead of $\tilde{\varphi}_{n,i}$ we obtain just
some linear combinations $\tilde{u}_{n}$.

\section{Examples}

\label{sec:examples}

Firoozi et al\cite{FMKJ22} applied the approach to one-particle Hamiltonian
operators of the form
\begin{equation}
H=-\frac{1}{2}\nabla ^{2}+V(x,y,z),  \label{eq:H_3D_general}
\end{equation}
where, for simplicity, we have resorted to a suitable dimensionless form\cite
{F20}. They proposed an improvement in the case of rotationally-invariant
potentials $V(x,y,z)=V(\rho ,z)$, $\rho =\sqrt{x^{2}+y^{2}}$, and chose $%
\psi (\mathbf{r},\tau )=U(\rho ,z,\tau )e^{im\phi }$, $m=0,\pm 1,\ldots $.
Therefore,
\begin{equation}
\frac{\partial U(\rho ,z,\tau )}{\partial \tau }=\left[ -\frac{1}{2\rho }%
\frac{\partial }{\partial \rho }\rho \frac{\partial }{\partial \rho }+\frac{%
m^{2}}{2\rho ^{2}}-\frac{1}{2}\frac{\partial ^{2}}{\partial z^{2}}+V(\rho
,z)\right] U(\rho ,z,\tau ).  \label{eq:Schro_time_dep_cylindrical}
\end{equation}
It is quite obvious that in this way one gets rid of the degeneracy of the
problems with this kind of symmetry. However, they applied this approach to
spherically symmetric models $V(x,y,z)=V(r)$, $r=\sqrt{x^{2}+y^{2}+z^{2}}$,
and, consequently, they failed to remove the degeneracy completely. For
example, in the case of the harmonic oscillator $V(r)=\frac{1}{2}r^{2}$ the
energy levels are
\begin{equation}
E_{n}=n+\frac{3}{2},\;n=n_{x}+n_{y}+n_{z},\;n_{x},n_{y},n_{z}=0,1,\ldots
,\;g_{n}=\frac{(n+1)(n+2)}{2}.  \label{eq:E_n_HO_cartesian}
\end{equation}
For the ground state $g_{0}=1$ and they obtained the correct wavefunction
that depends only on $r$. However, for the first two excited energy levels
they showed only one eigenfunction for each of them when there are $g_{1}=3$
and $g_{2}=6$.

For example, for $n=1$we have the following eigenfunctions
\begin{eqnarray}
\varphi _{1,\pm 1} &=&\frac{re^{-r^{2}/2}\sin {\left( \theta \right) }}{\pi
^{\frac{3}{4}}}e^{\pm i\phi },  \nonumber \\
\varphi _{1,0} &=&\frac{\sqrt{2}re^{-r^{2}/2}\cos {\left( \theta \right) }}{%
\pi ^{\frac{3}{4}}}.  \label{eq:varphi_1}
\end{eqnarray}
Note that, even when removing the dependence on $\phi $, they still depend
on the other angle $\theta $. Surprisingly, Firoozi et al\cite{FMKJ22}
plotted wavefunction vs $R\,(nm)$ in their figure 2. If we assume that $R$
stands for the radial variable $r$ there is something else amiss because the
plot range is $0\leq R\leq 0.4$ when one of the calculations was carried out
with a mesh size of $0.4\,nm$. Besides, those authors suggest that their
calculation provides $U(\rho _{i},z_{j},\tau _{k})$ which casts doubts on
those plots of wavefunction vs $R$ because the eigenfunctions (\ref
{eq:varphi_1}) exhibit an angular dependence.

In order to improve the approach in the case of spherically-symmetric
potentials we may choose $\psi (x,y,z,\tau )=R_{l}(r,\tau )Y_{l}^{m}(\theta
,\phi )$, where $Y_{l}^{m}(\theta ,\phi )$ are the spherical harmonics and $%
l=0,1,\ldots $, $m=0,\pm 1,\ldots ,\pm l$ the rotational quantum numbers.
The radial factor $R_{l}(r,\tau )$ is a solution to the equation
\begin{equation}
\frac{\partial R_{l}(r,\tau )}{\partial \tau }=\left[ -\frac{1}{2r^{2}}\frac{%
d}{dr}r^{2}\frac{d}{dr}+\frac{l(l+1)}{2r^{2}}+V(r)\right] R_{l}(r,\tau ),
\label{eq:radial_time_dep}
\end{equation}
and we can apply the approach safely because there are no degenerate states
left.

\section{Conclusions}

\label{sec:conclusions}

Throughout this paper we have reviewed a recently proposed approach termed
Q-BOR-FDTD method\cite{FMKJ22} and developed the main equations in a clearer
and more rigorous way. It has been shown that, although it provides the
eigenvalues, it fails to yield the eigenfunctions in the case of degenerate
states. This shortcoming affects the application of the method to the
spherically-symmetric potentials exhibited by some of the examples chosen by
the authors. For this reason, one may reasonably cast doubts on the validity
of some of their results. We have shown how to overcome this difficulty by
means of a simple and straightforward strategy that generalizes and improves
the one proposed by the authors.

\end{document}